# ABOUT THE SURVEY OF PROPAGANDISTIC MESSAGES IN CONTEMPORARY SOCIAL MEDIA

Yuri Monakhov[1], Natalia Kostina[1], Maria Medvednikova[1], Makarov Oleg[1], Irina Semenova[1]

[1]Department of Informatics and Information security, Vladimir State University, Vladimir 600000, Russia

This paper presents the research results that have identified a set of characteristic parameters of propagandistic messages. Later these parameters can be used in the algorithm creating special user-oriented propagandistic messages to improve distribution and assimilation of information by users.

## INTRODUCTION

Currently social networks are very popular, rapidly developing and becoming an integral part of the Internet users' lives.

The purpose of this research is to define the parameters of such promotional messages that influence the perception of information.

The object of research is social networking services on the Internet.

The subject under research is the process of promoting social networking and promotional communications options.

## FORMAL PARAMETERS OF PROPAGANDISTIC MESSAGES

During the research a number of propagandistic messages have been studied to identify the properties and parameters of those messages. Psychological experiments have been carried out, revealing the degree of significance of various parameters.

Messages for analysis were taken from the social networking service VKontakte, as the most advanced and popular in Russia.

As a result of the study the following options were designed.

1) The volume of messages. Measured by the number of symbols.
2) Message format. Calculated as the confidence to a specific format – text, video, image, sound.
3) The author. Determined by the degree of confidence in the author.
4) Relevance. Interval between the publications of official propaganda and messages related to a topic.
5) Targeting. Expresses the overlap of propagandistic message with the information of a user profile.
6) Transparency. Calculated as a percentage of common words in the message.
7) Identity. Percentage of words matches propagandistic messages with words from the user's messages.
8) Emotional color. Percentage of emotive words in the message (excluding stop-words).
9) "Nausea". The indicator of keywords repetition in the message. Sometimes it is called synonymous by density.
10) "Water". Percentage of insignificant words and stop-words in the message.

The psychological test to determine the impact of message formats on the perception of information was conducted. A message was represented in the form of video, audio, text and photo. Have been measured such parameters as the assimilation of the message content and emotional impact.

The experiment results showed that audio format is assimilated better than any other format.

Assessment of parameters was held by specific software. Were used such programs as Textus Pro, Advego, ZnakoSchitalka, Mira Tools.

The results of measurements in the programs were analyzed. Some patterns and matches the parameters in different programs were revealed.

In addition, there were proposed two approaches to the description of the dissemination of propagandistic messages: classical and non-classical.

The first approach assumes that the user can send propaganda both in private messages and on the "wall" of another user. It is possible to perform 2n actions, where n is less than the number of user's friends. An important factor is whether the user will send messages that have been posted to his "wall". To assess this factor, a survey was conducted. 70% of respondents gave the negative answer.

The second approach assumes that the user places a propagandistic message to the news line. In that case, he would not take any action to disseminate information. But then the message can be displayed in the news not all the friends of the site with probability

$$\int_{Tr_{ij}}^{1} P(Tr_{ij})dTr_{ij} = \gamma ,$$

where γ - the number of percent selected by the user.

Further study of the factors influencing the efficiency of distribution, has resulted in appearance of two categories of factors: technical and behavioral.

The technical factors include:
1) Variability of the message (different descriptions to the link in the message).
2) Recipient status (online / offline).
3) The age of the target audience (young people follow the new links more often.)

The behavioral factors include:
1) Friendly relations with a source - a message sent by a friend is more likely to be noticed than a message sent by a stranger.
2) Interest in the group - sending that uses targeting (the target audience of interest, university and so on) is more effective.
3) Filling the profile and attractiveness of a source and a group – filled profile of a source and complete information of a group inspires more confidence to the recipient.
4) Newsmaker - relevance of the message in a lucrative destination for the time interval increases the efficiency of distribution.
5) Age and sex of a source – for example, the messages, sent by the girls, aged 18 to 24, are more popular.
6) Activity in the group – the group should not be "dead": the more actual the information, the more effective delivery.
7) Geographical location of a source – some users have more confidence in groups and sources, located in the own country/region/city.

To quantify all of the parameters there were held the series of experiments with mass mailing messages users of the social networking service. After changing the parameters the results of review, the perception and response to messages have been changed as well.

## CONCLUSION

The result of this work can be considered as the development of a number of parameters that influence the effectiveness of advertising and propaganda in social networking services. This information may be used to increase resistance to propaganda, or, vice versa, to increase the digestibility of such messages.